\documentstyle[epsf]{elsart}
\begin{document}
\begin{frontmatter}
\title{Universal Susceptibility Variations\\ in 1+1 Dimensional Vortex Glass}
\author[Rutgers]{Chen Zeng}, 
\author[Rutgers]{P.L. Leath}, and 
\author[UCSD]{Terence Hwa}   
\address[Rutgers]{Department of Physics, Rutgers University,
Piscataway, NJ 08854, USA} 
\address[UCSD]{Physics Department, UCSD,  
La Jolla, CA 92093, USA}
\begin{abstract}
We model a planar array of fluxlines as a discrete solid-on-solid
model with quenched disorder. Simulations at finite temperatures
are made possible by a new algorithm which circumvents the slow
glassy dynamics encountered by traditional Metropolis Monte Carlo
algorithms. Numerical results on magnetic susceptibility 
variations support analytic predictions.  
\end{abstract}
%\pacs{PACS number: 74.60.Ge, 64.70.Pf, 02.60.Pn}
\end{frontmatter}

\newcommand{\br}{{\bf r}}
\newcommand{\bq}{{\bf q}}
\newcommand{\hL}{\widehat{\cal L}}

%\begin{twocolumn}

%%%%%%%%%%%%%%%%%%%%%%%%%%%%%%%%%%%%%%%%%%%%%%%%%%%%%%%%%%%%%%%%%%%%%%%%%%%%%%
%\paragraph*{Introduction}
%%%%%%%%%%%%%%%%%%%%%%%%%%%%%%%%%%%%%%%%%%%%%%%%%%%%%%%%%%%%%%%%%%%%%%%%%%%%%%

Over the years various low-temperature glass phases have been proposed
for vortex matter in dirty type-II superconductors\cite{blatter}.
Understanding the thermodynamics of such interacting many-body systems 
in the presence of quenched disorder remains a challenge. Similarities 
between the randomly-pinned vortex system and the more familiar
mesoscopic  electronic systems~\cite{meso} are highlighted by a recent 
experiment on vortices threaded through a thin 
crystal of $2$H-NbSe$_2$  by Bolle {\it et al.}~\cite{bolle}. 
Sample-dependent magnetic responses known as ``magnetic finger prints'' 
have been observed for such a mesoscopic vortex system.

%Because of the absence of topological defects such as dislocations
The simplicity of the planar vortex system allowed detailed 
theoretical studies of this glassy system~\cite{rough,mpaf,natt,gld,hnv,usf},
with quantitative predictions including 
the universal  sample-to-sample variation of
magnetic responses. However, until the work of Bolle {\it et al.}, 
there were hardly any experimental studies of this system, with 
difficulties stemming partly from the weak magnetic signals in 
such 2d systems. Meanwhile, progress on numerical approaches were 
also seriously hampered by the slow glassy dynamics often encountered 
in simulations at finite temperatures~\cite{2d_finite}. To overcome 
the this numerical difficulty, we model a planar array of
fluxlines as a discrete solid-on-solid model, viz., the dimer model
with quenched disorder. A new polynomial algorithm circumventing the
glassy dynamics enables us to simulate large systems of this
discrete model at any finite temperature.

%%%%%%%%%%%%%%%%%%%%%%%%%%%%%%%%%%%%%%%%%%%%%%%%%%%%%%%%%%%%%%%%%%%%%%%%%%%%%%
\paragraph*{The Model:}
%%%%%%%%%%%%%%%%%%%%%%%%%%%%%%%%%%%%%%%%%%%%%%%%%%%%%%%%%%%%%%%%%%%%%%%%%%%%%%
The dimer model consists of all complete dimer coverings $\{D\}$ on a
square lattice $\cal L$ as illustrated in Fig.~1(a). The partition
function is   
\begin{equation} 
Z=\sum_{\{D\}} \exp\left[-\sum_{<ij>\in D} \epsilon_{ij}/T_d\right]
\;\; , 
\label{eq1}
\end{equation} 
where the sum in the exponential 
is over all dimers of a given covering, and $T_d$ is the dimer temperature.
Quenched disorder is introduced
via random bond energies $\epsilon_{ij}$, chosen independently and
uniformly in the interval $(-\frac12,\frac12)$. 

\begin{figure}
  \begin{center}
    \leavevmode
    \epsfxsize=8cm
    \epsffile{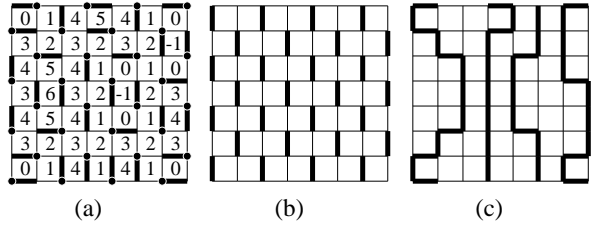}
  \end{center}
\caption{
(a) Snapshot of a dimer covering (thick bonds) together 
with the associated ``height'' values $h(\br)$
 for a lattice of size $L=8$ at temperature $T_d=1.0$. 
(b) Dimer covering of the fixed reference. (c) Vortex line configuration
(thick lines) obtained as the difference between (a) and (b).    
}
\label{fig_1}
\end{figure}

The dimer model is related to the planar vortex-line array via
the well-known mapping to the solid-on-solid (SOS) model (see Figs.~1):
Define integer heights $\{h\}$ at
the centers of squares of ${\cal L}$ which form the dual square
lattice ${\cal L}^D$, then orient all bonds of ${\cal L}^D$ such
that elementary squares of ${\cal L}^D$ that enclose sites of a
chosen sublattice of ${\cal L}$ (solid dots in Fig.1(a)) are circled
counterclockwise; Assign $-3$ to the difference of neighboring heights
along the oriented bonds if a dimer is crossed and $+1$ otherwise.
This yields single-valued heights up to an overall constant.
In terms of the height configuration $h(\br)$, the partition function 
(\ref{eq1}) can be written alternatively
as $Z = \sum_{\{h(\br)\}}
e^{-\beta {\cal H}[h]}$, where the SOS Hamiltonian
takes the following form in the continuum limit,
\begin{equation} 
\beta {\cal H}=\int d^2 \br\left[
\frac{K}{2} (\nabla h)^2
-{\bf f}({\bf r})\cdot\nabla h
+g\cos\left(G h \right)
\right]
\;\; .
\label{eq2}
\end{equation} 
Here, $K$ is an effective stiffness caused by the
inability of a tilted surface to take as much advantage of the 
low weight bonds as a flatter surface, and 
${\bf f}({\bf r})$ is a random local tilt bias. 
The periodicity of the cosine potential in (\ref{eq2}) is given by $G=2\pi/4$
since the smallest ``step'' of this height profile is four.
In the present context of a randomly pinned vortex array,
$h(\br)$ describes the coarse-grained 
displacement field of the vortex array with respect to 
a uniform reference state at the same vortex line density; see Fig.~1(c)
and Refs.~\cite{mpaf,hnv}.

%%%%%%%%%%%%%%%%%%%%%%%%%%%%%%%%%%%%%%%%%%%%%%%%%%%%%%%%%%%%%%%%%%%%%%%%%%%%%%
\paragraph*{The Algorithm:}
%%%%%%%%%%%%%%%%%%%%%%%%%%%%%%%%%%%%%%%%%%%%%%%%%%%%%%%%%%%%%%%%%%%%%%%%%%%%%%
Computing the partition function of complete dimer coverings on a weighted 
{\it planar} lattice $\cal G$ can be achieved in {\em polynomial} 
time~\cite{dimer_pfaff}. Weights here refer to the Boltzmann factors 
$w_{ij} \equiv \exp(-\epsilon_{ij}/T_d)$ on the bonds. In this article, 
we shall use the shuffling algorithm by Propp and 
coworkers\cite{dimer_shuffle} to sample the configurational space. 
The algorithm relates the partition function $Z_L$ on a lattice of 
linear size $L$ to $Z_{L-1}$ as $Z_L(\{w\})=C(\{w\}) Z_{L-1}(\{w^\prime\})$ 
after a simple weight transformation $\{w\}\rightarrow\{w^\prime\}$. 
The prefactor $C(\{w\})$ is independent of dimer coverings. 
The partition function is obtained in a ``deflation'' process in which
the above recursive procedure is carried out down to $L=0$ with
$Z_0=1$. This deflation process can also be reversed in an
``inflation'' process where a dimer covering at size $L-1$ can be used
to {\it stochastically} generate a dimer covering at size $L$ according to 
$Z_L$ already obtained. Repeating the inflation process thus generates
{\it uncorrelated} ``importance samplings'' of the dimer configurations, or
equivalently the equilibrium height configurations, without the need to 
run the slow relaxational dynamics. The ensuing numerical results are 
obtained by taking various measurements of the height configurations 
generated this way. A somewhat inconvenient feature of this approach is 
that the algorithm requires open boundary condition on the dimer model; 
this in turn fixes the total number of vortex lines, e.g.,  to $L/2$ 
on a $L\times L$ lattice (see Fig.~1).

\paragraph*{Numerical results:} 
Since the temperature of the vortex array, given by $K^{-1}$ in (\ref{eq2}), 
is generally different from the dimer temperature $T_d$, we first need to 
calibrate the temperature scale. To do so, we exploit a statistical 
rotational symmetry~\cite{symmetry,usf} of the system (\ref{eq2}) 
which guarantees that at large scales, the effective $K$ 
is the same as that of the pure system. In particular, $K$ can be obtained
by measuring the disorder-averaged thermal fluctuation of the displacement 
field $\overline{W_T^2}\equiv \overline{
\langle\prec h_T^2({\bf r})\succ - \prec h_T({\bf r})\succ^2\rangle}
= (2\pi K)^{-1} \ln(L)$, where $h_T({\bf r})
\equiv h({\bf r}) - \langle h({\bf r})\rangle$ measures the thermal distortion
superposed on the distorted ``background'' $\langle h({\bf r}) 
\rangle$ by the quenched disorder. We used $\prec{...}\succ$,
$\langle{...}\rangle$, and overline to denote spatial, thermal, and 
disorder averages respectively. With the polynomial algorithm, 
we were able to perform thorough disorder averages for equilibrated 
systems of sizes up to $512 \times 512$. To reduce boundary effects, 
we focus on the central $L/2\times L/2$ piece of the system
and compute its displacement fluctuations.
Fig.~2(a) illustrates the dependence of $\overline{W_T^2}(L)$ for
various dimer temperatures
$T_d$. The linear dependence on $\ln(L)$ is apparent. Identifying the
proportionality constant with $(2\pi K)^{-1}$, we obtain the empirical
relation between $K^{-1}$ and $T_d$ shown in Fig.~2(b). 
Note that our result 
recovers the exact relation $K^{-1}(T_d\to\infty) = 16/\pi$ for  
the dimer model without disorder~\cite{exact}.
Since the glass transition of the system (\ref{eq2}) is expected to
occur at temperature
$K_g^{-1} =4\pi/G^2 = 16/\pi$, our system is glassy for the
entire range of dimer temperatures.

\begin{figure}
  \begin{center}
    \leavevmode
	\epsfxsize=8.5cm
    \epsfysize = 1.5in
    \epsffile{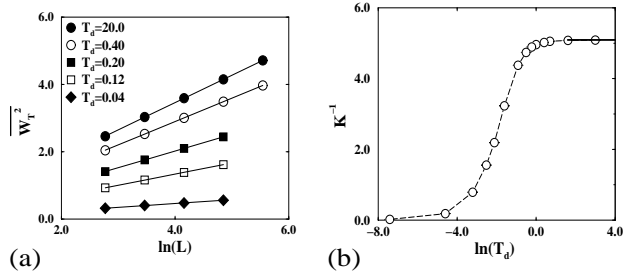}
  \end{center}
\caption{(a) The disorder-averaged thermal displacement fluctuation 
$\overline{W_T^2}(L)$ 
for various dimer temperatures $T_d$. Each data point was obtained by 
generating $10^3$ thermal samplings for each of $10^3$ 
disorder realizations. The straight lines are fit to 
$\overline{W_T^2} = (2\pi K)^{-1} \ln(L) + {\rm const}$.
 (b) The extracted relation between  $K^{-1}$ and $T_d$. 
The horizontal line indicates the asymptotic value of $K^{-1}$
as $T_d \to\infty$.
}
\label{fig_2}
\end{figure}

To probe this glassy state further, we now focus on its 
magnetic response, i.e., the uniform magnetic susceptibility
$\chi$. Due to the constraint of fixed vortex density when 
using the dimer representation, it is not easy to compute the 
magnetic susceptibility directly by varying the vortex fugacity
(i.e., external magnetic field).
Instead, we use the fluctuation-dissipation relation,
\begin{equation}
\chi \propto 
\lim_{q\rightarrow 0} \bq^2
\left(
\langle \hat{h}_{\bq} \hat{h}_{-\bq} \rangle 
-\langle \hat{h}_{\bq}\rangle \langle \hat{h}_{-\bq}\rangle
\right)
\equiv \lim_{q\rightarrow 0} \bq^2 C_2(\bq) , 
\label{eq3}
\end{equation}
%\begin{eqnarray} 
%\chi &\propto& 
%\int\int d{\br}_1 d{\br}_2 
%\left( \langle \partial_x h(\br_1) \partial_x h(\br_2) \rangle
%- \langle \partial_x h(\br_1)\rangle
%  \langle \partial_x h(\br_2)\rangle 
%\right) \cr
%&=& 
%\lim_{q\rightarrow 0} \bq^2
%\left(
%\langle \hat{h}_{\bq} \hat{h}_{-\bq} \rangle 
%-\langle \hat{h}_{\bq}\rangle \langle \hat{h}_{-\bq}\rangle
%\right)
%\equiv \lim_{q\rightarrow 0} \bq^2 C_2(\bq) , 
%\label{eq3}
%\end{eqnarray}
where $\hat{h}_{\bq}$ is the Fourier transform of $h(\br)$.  
$\chi$ only probes the system at the largest scale as indicated 
by the $q\rightarrow 0$ limit, which 
is evaluated at $(q_x,q_y)=(\pi/L,0)$ for finite systems. 

As shown in Figs.~3, the distribution $P(\chi)$ at low temperature
is peaked at zero and decays more strongly than any power laws  
for large $\chi$'s. This indicates that the glass phase is 
typically {\em rigid} against
the penetration of additional fluxlines when 
the external magnetic field is increased by a small amount.
The mean value $\overline{\chi}$, which takes on a finite 
value $\propto K^{-1}$ due to the statistical rotational 
symmetry~\cite{usf,symmetry}, 
is however controlled by the rare events contained in
the  tail of $P(\chi)$. This is similar to the behavior found
for a single flux line in $1+1$ dimensional random 
potential~\cite{mezard,hf}.

\begin{figure}
  \begin{center}
    \leavevmode
        \epsfxsize=8.5cm
    \epsfysize = 1.5in
    \epsffile{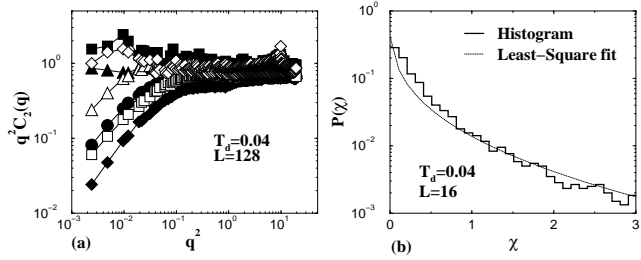}
  \end{center}
\caption{(a) $q^2 C_2(\bq)$ vs.~$q^2$. Different symbols denote  
different disorder realizations, all of which are for
$L=128$ at $T_d=0.04$. The susceptibility $\chi$ is evaluated
at $q=\pi/L$. (b) Probability distribution $P(\chi)$
computed from $6\times 10^3$ disorder realizations for $L=16$ at
$T_d=0.04$. Dotted line is a fit to a stretched exponential
$P(\chi) = A \exp (-B \chi^{\alpha})$ with $\alpha = 0.5(2)$.  
}
\label{fig_3}
\end{figure}
 
\begin{figure}
  \begin{center}
    \leavevmode
	\epsfxsize=8.5cm
    \epsfysize = 1.9in
    \epsffile{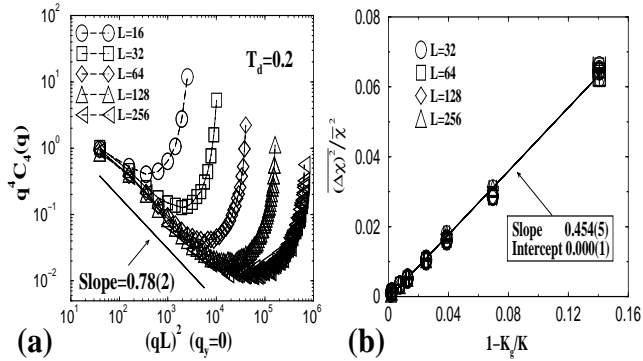}
  \end{center}
\caption{(a) $q^4 D(\bq)$ vs.~$(qL)^2$ along $q_y=0$ axis.
Data are obtained by averaging over $10^3$ disorder realizations 
for $L=16,32,64,128,256$ at $T_d=0.2$. The solid line is 
the least square fit; (b) Fractional variance as a function 
of the reduced temperature, $\tau=1-K_g/K$, for $L=32, 64, 128$, and 
$256$ averaged over $10^3$ disorder realizations.
The solid line is a least-square linear fit for $|\tau| < 0.15$.
}
\label{fig_4}
\end{figure}

A striking feature of this vortex glass 
concerns the sample-to-sample variation of the 
susceptibility $\chi$. It was predicted~\cite{usf} 
that the fractional variance,
\begin{equation}
{\rm var}[\chi]/\overline{\chi}^2 \stackrel{L\to\infty}{\rightarrow} D |\tau|
\qquad {\rm for} \quad 0 < -\tau \ll 1,
\label{Fsus}
\end{equation}
is {\em universal},
with $D$ being a computable, size-independent constant of order unity.
Using Eq.~(\ref{eq3}), we obtain 
${\rm var}[\chi] \propto \lim_{\bq\rightarrow 0} \bq^4 C_4(\bq)$
where $C_4(\bq)\equiv \overline{C_2^2(\bq)}-{\overline{C_2(\bq)}}^2$.
%\begin{equation}
%{\rm var}[\chi] \propto \lim_{\bq\rightarrow 0} \bq^4 
%\left(
%\overline{C_2^2(\bq)}-{\overline{C_2(\bq)}}^2
%\right) 
%\equiv \lim_{\bq\rightarrow 0} \bq^4C_4(\bq) \;\; .  
%\end{equation} 
As shown in Fig.~\ref{fig_4}(a), $\bq^4C_4(\bq)$ obeys a power-law 
scaling as $\bq^4C_4(\bq) \sim (qL)^{0.78(2)}$ for small $q$.
This behavior is consistent with the analytical prediction that 
the variance is size-independent because the 
$q\rightarrow 0$ limit is evaluated at $(q_x,q_y)=(\pi/L,0)$
for finite systems. The power-law scaling and its exponent are, however,  
not understood at present. Finally the temperature
dependence of the fractional variance, as shown in 
Fig.~\ref{fig_4}(b), can  be
deduced from the relation between $K^{-1}$ and $T_d$ given in Fig.~2(b). It is
well described by the linear form (\ref{Fsus})
close to the glass transition with $D= 0.454(5)$. 

%%%%%%%%%%%%%%%%%%%%%%%%%%%%%%%%%%%%%%%%%%%%%%%%%%%%%%%%%%%%%%%%%%%%%%%%%%%%%%
\paragraph*{Conclusion.}
%%%%%%%%%%%%%%%%%%%%%%%%%%%%%%%%%%%%%%%%%%%%%%%%%%%%%%%%%%%%%%%%%%%%%%%%%%%%%%
Finite temperature simulations were performed on 
a disordered dimer model to study the magnetic response of a planar 
vortex array. Universal susceptibility variations in 
the collective pinning regime were observed;  
critical behavior near the glass transition 
compared well with analytic predictions. 
It is hoped that the present study will stimulate further experimental
investigations of the physics of mesoscopic vortex systems.

%\end{twocolumn}

\begin{thebibliography}{99}

\bibitem{blatter}
G. Blatter {\it et al},
Rev. Mod. Phys. {\bf 66}, 1125 (1994).

\bibitem{meso}
E. Akkermans {\it et al} eds.,
{\it Mesoscopic Quantum Physics}, (North Holland, Amsterdam, 1994).


\bibitem{bolle} 
C.A. Bolle {\it et al.}, Nature {\bf 399}, 43 (1999).

\bibitem{rough}
See, e.g. J.L. Cardy and S. Ostlund, 
Phys. Rev. B {\bf 25}, 6899 (1982);
J. Toner and D. P. DiVincenzo,
Phys. Rev. B {\bf 41}, 632 (1990);

\bibitem{mpaf}
M.P. Fisher, 
Phys. Rev. Lett. {\bf 62}, 1415 (1989).  

\bibitem{natt}
T. Nattermann, I. Lyuksyutov, and M. Schwartz, Europhys. Lett. {\bf 16},
295 (1991);

\bibitem{gld}
T. Giamarchi and P. Le Doussal,
Phys. Rev. B {\bf 52}, 1242 (1995). 

\bibitem{hnv}
T.~Hwa, D.R.~Nelson, V.~M.~Vinokur, Phys.~Rev.~B {\bf 67}  1167 (1993).


\bibitem{usf}
T. Hwa and D.S. Fisher, 
Phys. Rev. Lett. 72, 2466 (1994). 

\bibitem{2d_finite} 
G. G. Batrouni and T. Hwa,
Phys. Rev. Lett. {\bf 72}, 4133 (1994); 
E. Marinari, R. Monasson, and J. J. Ruiz-Lorenzo,
J. Phys. A {\bf 28}, 3975 (1995); 
D. Cule and Y. Shapir,
Phys. Rev. Lett. {\bf 74}, 114 (1995); 


\bibitem{dimer_NP}
L.G. Valiant, 
Theo.~Computer Science {\bf 8}, 189 (1979).  

\bibitem{dimer_pfaff}
P.W. Kasteleyn,
Physica {\bf 27}, 1209 (1961); 
M.E. Fisher, 
J. Math. Phys.  {\bf 7}, 1776 (1966). 

\bibitem{dimer_shuffle}
N. Elkies, G. Kuperberg, M. Larsen, and J. Propp, 
J. Algebraic Comb. {\bf 1}, 111 \& 219 (1992); 
J. Propp, ``Urban renewal'', available  
from http://www-math.mit.edu/$\sim$propp/articles.html. 

\bibitem{exact}
See, e.g., R.W. Youngblood, D.J. Axe, and B.M. McCoy, 
Phys. Rev. B {\bf 21}, 5212 (1980); C.L. Henley, 
J. Stat. Phys. {\bf 89}, 483 (1997).  

\bibitem{symmetry}
U. Schulz {\sl et al}, 
J. Stat. Phys. {\bf 51}, 1 (1988). 

\bibitem{mezard}
M. Mezard, J. Phys. (Paris) {\bf 51}, 1831 (1990).

\bibitem{hf}
T. Hwa and D.S. Fisher, Phys. Rev. B {\bf 49}, 3136 (1994). 
\end{thebibliography}
\end{document}